# Cluster structure of superconducting phase and the nature of peaks in the doping dependences of the London penetration depth in iron pnictides

K.V. Mitsen[*], O.M. Ivanenko

Ginzburg Center for High-Temperature Superconductivity and Quantum Materials, Lebedev Physical Institute, Moscow, 119991, Russia

ABSTRACT

The mechanism of formation of dome-like phase diagrams and the features of magnetic field penetration in iron pnictides with hetero- and isovalent doping are considered within the framework of the previously proposed model, which assumes the local character of doping and the cluster structure of resulted superconducting phase. It is shown that the proposed model, despite its simplicity and neglect of the features of the electronic structure, makes it possible not only to accurately calculate the positions of superconducting domes on the phase diagrams of specific HTSC compounds, but also to explain the nature and position of sharp peaks in the London penetration depth depending on the doping level.

**Introduction**

Nanoscale electronic inhomogeneity is a long-standing unsolved problem in the physics of HTSC compounds [1,2]. One of the possible reasons for this inhomogeneity, is assumed to be the statistical nature of doping [3].

Earlier in [4], we showed that the features of the phase diagrams of heterovalently doped cuprate and pnictide HTSCs can be understood within the framework of an approach that assumes the self-localization of doped carriers. At such approach doped carrier locally deforms the electronic structure of the crystal, leading to the formation of so-called CT plaquettes in the cells adjacent to the localization area of the doped carrier ($CuO_4$ cells in cuprates and $AsFe_4$ cells in pnictides). Due to the influence of the dopant on these plaquettes, the gap $\Delta$ for the transitions between the states of the anion and the cation in this plaquette is reduced to the value $0<\Delta^*<E_{ex}$ - the binding energy of the CT exciton.

In the undoped phase, $\Delta>E_{ex}$, and excitons cannot exist. In the case of heterovalent doping, the doped electron (hole) is distributed in accordance with the lattice symmetry over the nearest 4 cations (anions), changing their charge by $\pm e/4$. The placement of $\pm e/4$ charge on the appropriate ion of the $CuO_4$ or $AsFe_4$ plaquette reduces $\Delta$ to a value $0<\Delta^*<E_{ex}$, which is sufficient for the exciton generation in this plaquette. The resulting excitons form a bound state (trionic complex) with a doped carrier, which is the reason for its self-localization.

The placement of two charges $\sim e/4$ on one plaquette can, depending on the value of $\Delta$, either vanishes the gap or reduces it to a value of $\Delta^{**}$. According to estimates [5], in the case of heterovalent doping in both cuprates and pnictides, the joint action of two doped carriers on one plaquette, inducing two charges $\sim e/4$ on it, leads to vanishing of the gap and corresponds to the transition of this plaquette to the "metallic" state.

As dopant concentration increases, CT plaquettes formed around the dopant are combined into CT clusters of various sizes, which form the CT phase. In such a cluster, electrons in the band formed by the orbitals of Fe or O ions interact resonantly with CT exciton states in CT plaquettes. For heterovalent doping, the doping range corresponding to the region of existence of CT clusters can be determined for each specific compound. As has been shown [4], these ranges coincide with good accuracy with the regions of superconducting domes in the phase diagrams of all studied cuprate and pnictide HTSCs with heterovalent doping. Based on this, the CT phase, which is either a percolation cluster of CT plaquettes or a network of Josephson-coupled CT clusters, was identified by us with the HTSC phase. A distinctive feature of CT cluster [4] is the possibility of the formation in them of Heitler-London centers - ionic complexes uniting the pairs of neighboring CT-plaquettes. The interaction of band electrons with Heitler-London centers, as shown, can lead to the appearance of free carriers in such a system and can be the very mechanism responsible for superconducting pairing [4,6].

---

[*] E-mail addresses: mitsen@lebedev.ru

As for isovalently doped pnictides, the mechanism of formation of CT plaquettes in them differs from that in the case of heterovalent doping. In this case, the change in the electronic structure around the dopant results from the difference between its intrinsic field and the field of the matrix ion. Nevertheless, as we will show below, in this case it is also possible to determine the concentration range of isovalent impurities corresponding to the existence of CT clusters, or which is the same, the region of the superconducting dome.

In this paper, we are going to show that, in addition to determining the positions of superconducting domes on the phase diagrams of cuprates and iron pnictides, the proposed model accurately traps subtle features in the behavior of their various characteristics, including sharp anomalies in doping dependences of the London penetration depth, resistance anisotropy and others. Taking into account that the interaction of itinerant electrons with excitonic states is genetically inherent in proposed model, it can be assumed that the exciton mechanism contributes to superconducting pairing in cuprates and pnictides.

**Heterovalent doping**

In contrast to high-Tc cuprates superconductivity in iron pnictides can be induced by partial substitution of Fe in the conducting layers by other transition metal elements like Co and Ni. Compared to $Fe^{2+}$ ion, $Co^{2+}$ ($3d^7$) has one more 3d electron, but $Ni^{2+}$ ($3d^8$) has two more 3d electrons. Then it is expected that each Ni dopant induces two extra electrons while each Co dopant only induces one extra electron [7].

According to [4], in Ba$(Fe_{1-x}Co_x)_2As_2$ the doped electron is distributed over the 4 nearest Fe ions (denoted in Fig. 1 as Fe+e/4), giving rise to CT- excitons in the surrounding cells and thus forming a belt of 8 CT plaquettes $AsFe_4$ (Fig. 1 a). The resulting excitons form a bound state with a doped electron (trionic complex), which prevents its further delocalization.

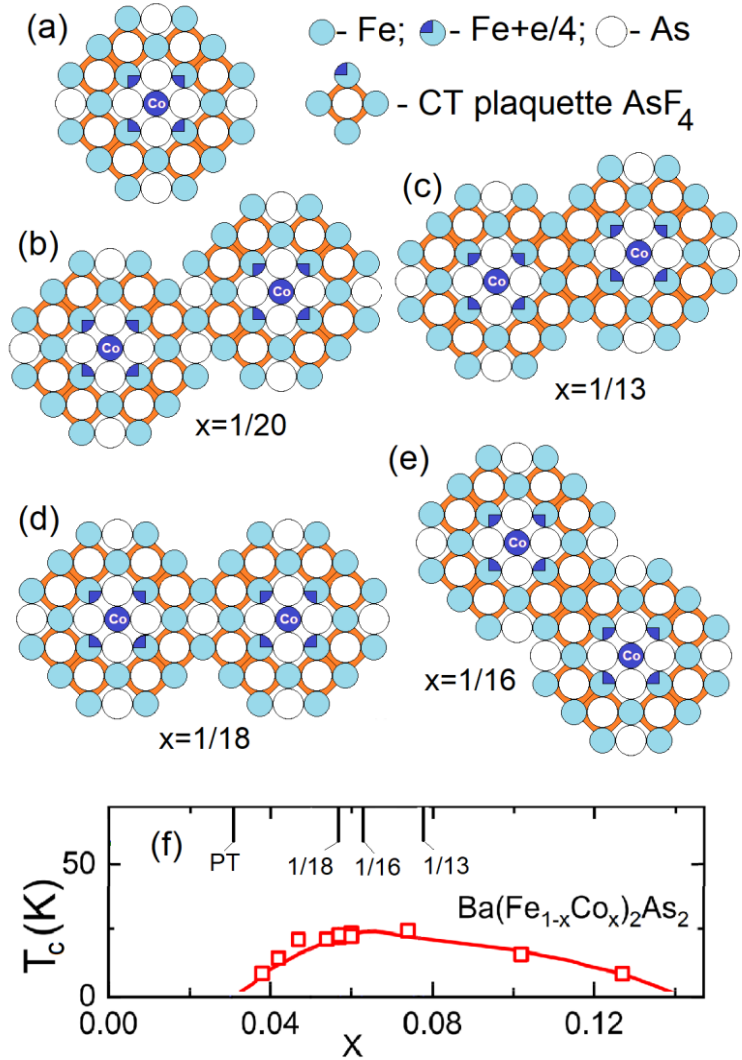

**Figure 1.** Formation of CT clusters in iron pnictides Ba122 with heterovalent Co doping. (**a**) at the substitution of a Fe ion with Co, the emerging doped electron is distributed over four adjacent Fe ions (denoted as Fe+e/4) to produce a belt of eight CT plaquettes. The doped electron and the CT plaquettes generated by it form a trionic complex; (**b**) the maximum distance between trionic complexes $l=\sqrt{20}$, at which the formation of a CT cluster is possible. It corresponds to a dopant concentration x = 1/20 at ordered distribution of trionic complexes; (**c**) the minimum distance between trionic complexes $l=\sqrt{13}$, at which the formation of a CT cluster is possible. It corresponds to a dopant concentration x = 1/13 at ordered distribution. At a higher concentration (lower $l$) the CT gap in the CT plaquettes will be vanished by the joint impact of two neighboring dopants. Thus the number of CT plaquettes, and, therefore, the volume of the CT phase will rapidly decrease; (**d,e**) intermediate dopant concentrations, providing the formation of large CT clusters (superclusters) due to the possibility of trionic complexes at these concentrations to form the sub-lattices retaining the corresponding symmetry elements of the crystal matrix; (**f**) superconducting phase diagram of Ba(Fe$_{1-x}$Co$_x$)$_2$As$_2$ [8]: PT – percolation threshold.

When neighbouring trionic complexes are located at a definite distance $l$ apart the formation of clusters of CT plaquettes is possible. In this case, trionic complexes can combine with each other into a cluster either by Fe sites (Fig. 1 b, d) or by Fe-Fe sides (Fig. 1 c, e). Let call the cluster with the parameter $l$ if it formed by trionic complexes located at the same distance $l$ from each other. The continuous cluster with the $l$ parameter can exist [9] only in the doping range from the threshold of site percolation on the lattice with the parameter $l$ to its complete filling:

$$0.593/l^2 < x < 1/l^2 \qquad (1)$$

As follows from [4], such a cluster is a cluster of the HTSC phase. Thus, it can be expected that an additional channel for conductivity opens along the percolation CT cluster, which should be accompanied by a sharp decrease in the resistance when the concentration of the dopant exceeds the percolation threshold. This conclusion is confirmed by experiment [10].

If we know the geometry of trionic complexes for each HTSC compound, we can determine the ranges of dopant concentrations corresponding to the existence of the percolation clusters of CT plaquettes with different $l$ parameters. From fig. 1b it can be seen that for $Ba(Fe_{1-x}Co_x)_2As_2$, the maximal $l$ value that ensures the formation of a CT cluster is $l=\sqrt{20}$ (in units of the lattice constant). Such clusters, according to (1), can be formed starting from x = 0.03 and reach the maximum density of CT plaquettes with an ordered arrangement of trionic complexes in a square sublattice at x = 0.05.

It is obvious that the formation of superclusters is facilitated by the arrangement of trionic complexes at the sites of a square lattice with a certain parameter $l$. What could be the mechanism of this ordering?

In fig. 1 we have depicted a trionic complex centered on Co dopant. This is done to show that the trionic complex is formed by a carrier that is introduced by Co dopant. If we assume that the Co dopants are rigidly bound to trion complexes, then the ordering of the latter into the sub-lattice should occur together with the ordering of the dopants. However, it can be assumed that trionic complexes can be detached from the dopant and move through the crystal as a whole, retaining their spatial structure. In this case, the doped charge can be likened to a polaron, which polarizes not the lattice, but the electronic structure around it.

Since trionic complexes carry a charge, with increasing concentration they can order into a sub-lattice with parameter $l$ in accordance with their concentration. This sub-lattice of trionic complexes will be pinned by an irregular lattice of dopants.

We have not found strong experimental confirmation for either of the two mechanisms. However, some experimental results, which we will discuss below, give us grounds to assume that in the case of heterovalent doping, ordering of trionic complexes occurs against the background of an irregular lattice of dopants. For now, we will consider here that such an ordering mechanism in heterovalent-doped pnictides exists.

As the concentration increases, CT plaquettes in the neighbouring trionic complexes start touching one another, which is accompanied with the formation of trionic clusters characterized by a certain value of $l$.

The smallest $l$ value for a cluster of CT plaquettes will be $l=\sqrt{13}$ (Fig. 1c). This distance between trionic complexes corresponds to their concentration x = 1/13≈0.077 at ordered arrangement in square lattice. At $l<\sqrt{13}$, some of the plaquettes are under the simultaneous action of two doped electrons, and the conditions for the appearance of excitons cease to be fulfilled in them. This leads to a rapid decrease in the concentration of CT plaquettes (and the volume of HTSC phase) with an increase in x> 0.077 and to their disappearance at $l=3$ (x≈0.11), when, with complete Co ordering, each plaquette is under the joint action of two dopants.

Thus, based on the proposed model, one should expect that in $Ba(Fe_{1-x}Co_x)_2As_2$ the concentration range corresponding to the existence of CT clusters is 0.03 <x <0.11, which is in good agreement with the region of the superconducting dome on phase diagram of this compound (Fig. 1 f) [8].

Note that in the range 1/20 <x <1/13, the formation of other ordered structures of trionic complexes is possible, providing the formation of a percolation CT cluster: at x = 1/18, 1/17, and 1/16. However, of all the possible types of sublattices into which trionic complexes can be ordered, only sublattices with $l = \sqrt{18}$ and 4 allow the formation of large clusters (with the number of trionic complexes >>1 in each cluster), since the corresponding symmetry elements of the crystal matrix should be retained in a large cluster of ordered trionic complexes. We will refer to such clusters of trionic complexes as superclusters.

These superclusters of ordered trionic complexes in the FeAs planes are separated by regions of the disordered phase and, in the superconducting state, are linked to each other by Josephson links. It can be assumed that at a Co concentration close to x = 1/18 and 1/16, this charge ordering can initiate a LTO-LTT transition in the crystal, similar to the LTO-LTT transition observed in $La_{2-x}Ba_xCuO_4$ at x≈1/8. [11]. This assumption finds its experimental confirmation in the work [12], the authors of which observed at T <25 K a sharp suppression of the Orthorhombic Lattice Distortion in single crystals $Ba(Fe_{1-x}Co_x)_2As_2$ with x> 0.057 and a transition to tetragonal symmetry at x = 0.062.

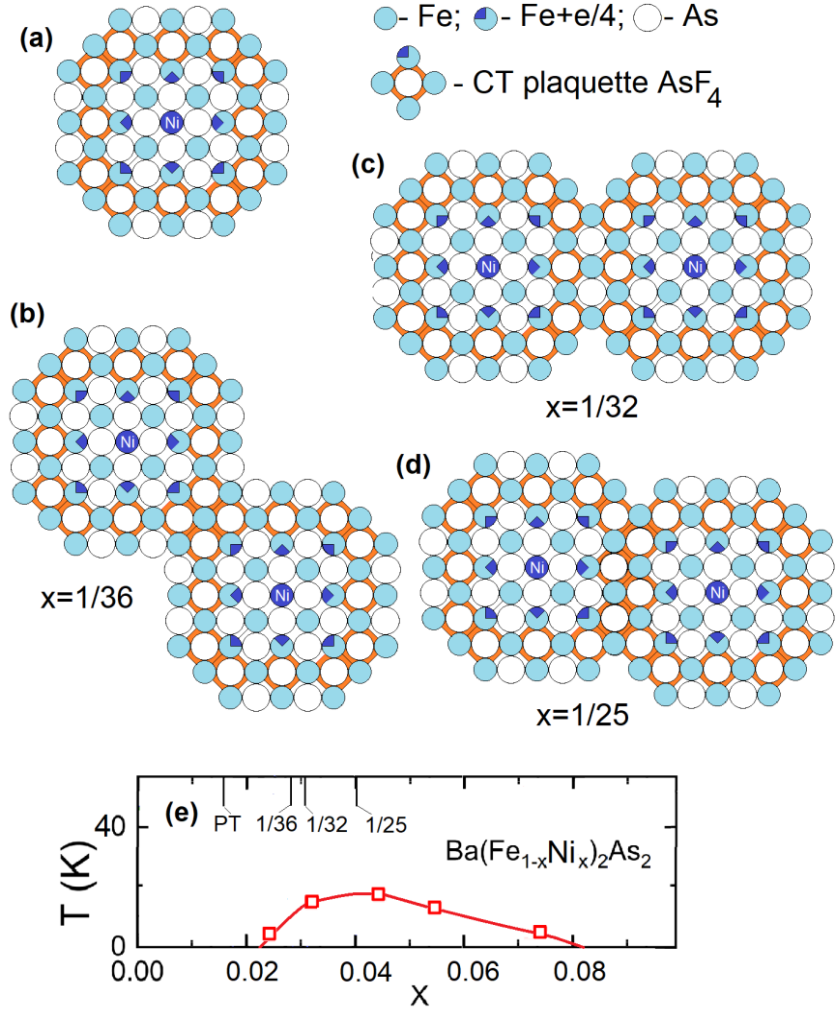

Fig. 2. Formation of CT clusters in iron pnictides Ba122 with heterovalent Ni doping. (a) at the substitution of a Fe ion with Ni, two doped electrons are distributed over eight Fe ions in the second coordination sphere (denoted as Fe+e/4) to produce a belt of twelve CT plaquettes; (b,c) the mutual arrangement of two trionic complexes at dopant concentrations, which allow the formation of large CT clusters (superclusters) due to the possibility of trionic complexes at these concentrations to form sub-lattices that retain the corresponding symmetry elements of the crystal matrix; (d) the optimal concentration for reaching of high density of CT plaquettes is x=1/25; (e) superconducting phase diagram of $Ba(Fe_{1-x}Ni_x)_2As_2$ [13]: PT – percolation threshold.

As noted above, every Ni donates two electrons into Fe layer. Accordingly, the region of their localization expands until the excess charge on 8 Fe ions in the second coordination sphere becomes equal to ~e/4, and the emerging trionic complex takes the shape shown in Fig. 2a. Since for the formation of a superclusters, it is necessary for the trionic complexes to be ordered into the sub-lattice retaining the corresponding symmetry elements of the crystal matrix, there are two concentration values x=1/32 and 1/36 when superclusters can be formed (Fig.2 b,c). At the same time, it is easy to see that the optimal concentration providing high density of CT plaquettes is x=1/25 (Fig. 2d). This concentration value corresponds to the maximum $T_c$ in the phase diagram (Fig. 2e) [13].

However, in contrast to Ba(Fe$_{1-x}$Co$_x$)$_2$As$_2$, where superclusters are formed in the region of optimal doping, in Ba(Fe$_{1-x}$Ni$_x$)$_2$As$_2$ such superclusters can exist in the underdoped region (in the vicinity of the range 1/36 <x <1/32). This conclusion is confirmed by the results of [14], where the restoration of resistance isotropy in a narrow region was observed (Fig. 3).

We associate the resistance anisotropy with proximity to the percolation threshold on trionic complexes. The point is that the transition from tetragonal to orthorhombic phase at a temperature $T_s$ results in a difference in the values of the percolation thresholds for different directions. As result, the percolation threshold is reached first in one of the directions at a concentration $x_p$. With a further increase in x>$x_p$, the power of the current-carrying part of percolation cluster providing percolation in one direction will be higher than in the other one, which reveals as the resistance anisotropy.

Obviously, the anisotropy should vanish when the trionic complexes fill the basal plane entirely. However, this is also possible at lower concentrations if superclusters of trionic complexes can be formed at these concentrations. This is what we believe to be the case in Ba(Fe$_{1-x}$Ni$_x$)$_2$As$_2$.

In accordance with the proposed consideration, the maximum anisotropy should be observed at the site percolation threshold on the lattice with $l$=6 (Fig. 2b). Corresponding value of $x_p$=0.593×(1/36)=0.016. Also it can be seen that the anisotropy practically disappears in the range 0.025 <x <0.028, i.e. where ordering of trionic complexes into regular symmetric lattices takes place. The difference from the calculated interval 0.028<x<0.031 is apparently due to the error in determining the composition. In the same range, with a decrease in temperature below Ts, a transition to the LTT phase should be expected, similar to Ba(Fe$_{1-x}$Co$_x$)$_2$As$_2$.

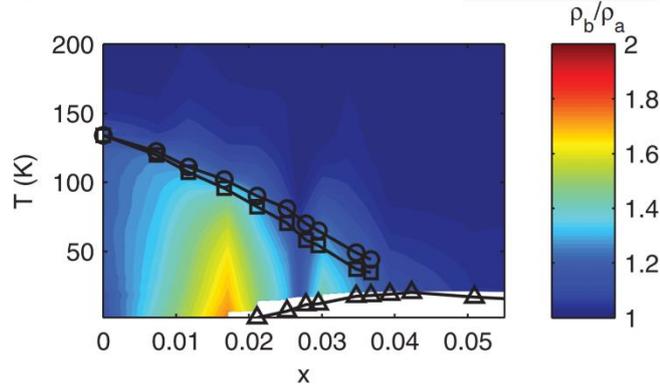

Fig. 3. In-plane resistivity anisotropy $\rho_b/\rho_a$ as a function of temperature and doping for Ba(Fe$_{1–x}$Ni$_x$)$_2$As$_2$. Black circles, squares, and triangles indicate $T_s$ (structural transition temperature), $T_N$ (Neel temperature), and $T_c$, respectively [14].

**Isovalent doping**

The dome-like shape of the $T_c(x)$ dependences observed both in cuprates and iron pnictides suggests a similarity of the mechanisms leading with doping to a transition of the system to a state with the same mechanism of superconducting pairing. It is surprising, however, that the same dome-like dependence of $T_c(x)$ is also observed for iron pnictides with an isovalent substitution of ions in the basal plane, e.g., in Ba(FeAs$_{1-x}$P$_x$)$_2$.

As previously noted, for isovalent doping a change of local electronic structure around the dopant is related to a distinction between its intrinsic field and the field of the substituted ion. In the case of isovalent doping, the role of an additional charge affecting the value of Δ plays a change of electron density near Fe cations, determined by the difference of ionic radii of the dopant and substituted ion. In contrast to heterovalent doping, where the charge q ~ ±e / 4 is induced directly on the plaquette and suppresses the gap for the transition between the central and all four external ions of the plaquette, with isovalent doping, the P dopant ion affects only the surrounding Fe ions.

Thus, the substitution of a $P^{3-}$ ion of a smaller ionic radius for an $As^{3-}$ ion is equivalent to a decrease of the negative charge near four adjacent Fe ions, which reduces the gap of $\Delta$ for the transfer of an electron from others As ions to these Fe cations to a value of $\Delta^*$ (Fig. 4a). Such Fe ions will be further called Fe′ ions. As this reduction of the gap $\Delta$ due to a difference of ionic radii is expected to be smaller than that under heterovalent doping, two ions of P can be positioned next to the Fe ion to reduce the gap additionally to a value $0<\Delta^{**}<\Delta^*$. Such Fe ions, whose neighbours are two P ions, will be designated as Fe″ (Fig. 4a).

The ground state of undoped and weakly doped $Ba(FeAs_{1-x}P_x)_2$ is SDW state. To form in FeAs plane CT plaquettes $AsFe'_4$ or $AsFe''_4$ (highlighted in brown), it is necessary that the As ion be surrounded by four Fe′ or Fe″ ions, correspondingly. For this reason, as seen in Fig. 4 b, at ordered arrangement of P ions with $x = 0.2$ no CT plaquettes are formed and SDW state takes place.

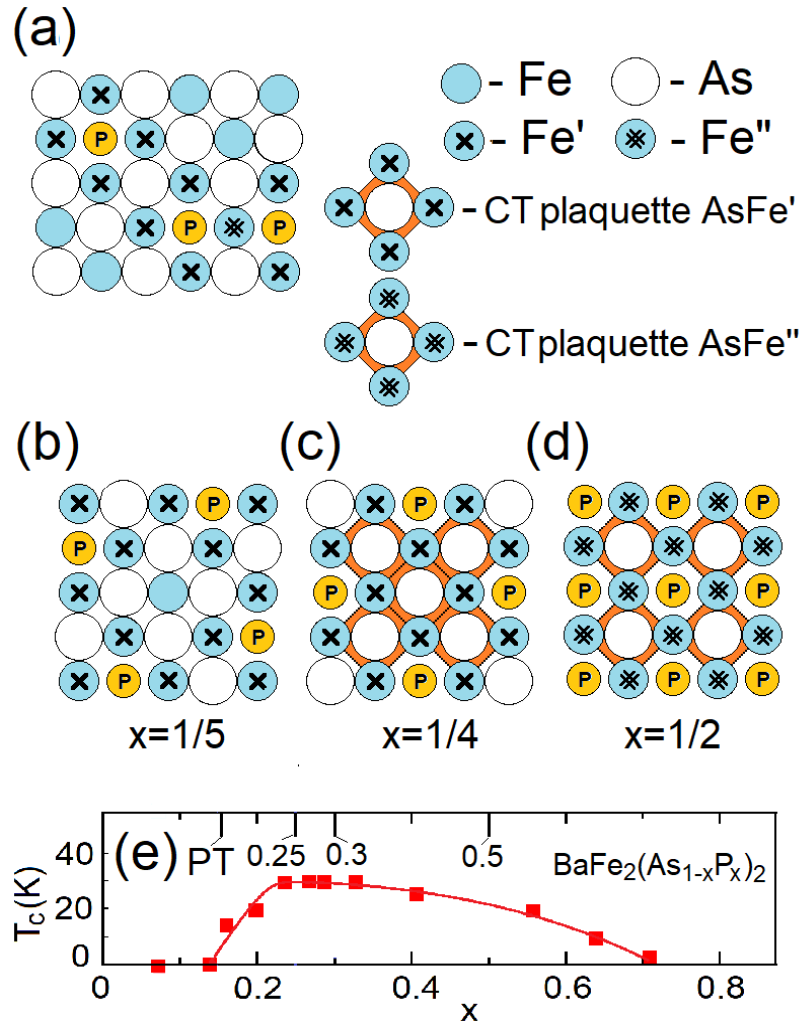

**Fig. 4 (a)** Substitution of $P^{3-}$ ion of a smaller radius for $As^{3-}$ is equivalent to a decrease of the negative charge near four adjacent Fe ions, which reduces the gap $\Delta$ for the transfer of an electron from As ions to these Fe cations to a value of $\Delta^*$; **(b)** at an ordering of P ions into a $\sqrt{5}\times\sqrt{5}$ lattice ($x = 0.2$), $AsFe'_4$ CT plaquettes are absent; **(c)** at $x > 0.15$, formation of CT clusters from $AsFe'_4$ plaquettes, which at $x = 0.25$ can occupy up to 75% of the basal plane, is possible; **(d)** at $x > 0.3$, the existence of CT clusters from $AsFe''_4$ plaquettes, which at $x = 0.5$ can occupy up to 50% of the basal plane, is possible; **(e)** the dependence of $T_c$ of $Ba(FeAs_{1-x}P_x)_2$ on the doping level [15]: PT – percolation threshold.

The concentration of CT plaquettes ($AsFe'_4$) increases with doping (as does $T_c$ of crystal) and reaches a maximum at ordered arrangement of dopants, corresponding to x = 0.25 (Fig. 4c). At this concentration value, the formation of superclusters (with a number of CT plaquettes included in each cluster >>1) is possible, since they retain the corresponding symmetry elements of the crystal matrix. This is confirmed by measurements of the in-plane resistivity anisotropy in Ba(FeAs$_{1-x}$P$_x$)$_2$ as a function of temperature and doping (Fig. 5) [16]. It is seen that the anisotropy practically disappears in a narrow region around x = 0.25.

At the same time, according to (1), the existence of small CT clusters with $l = 2$ is possible starting from the percolation threshold $x_p$=0.593x0.25≈0.15. Thus, in the range 0.15<x<0.25, SDW and HTSC phases will coexist on nanoscale.

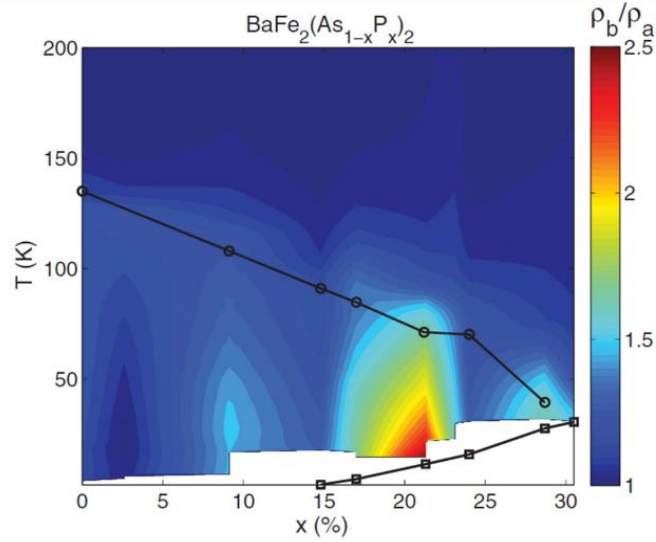

Fig. 5 In-plane resistivity anisotropy $\rho_b/\rho_a$ as a function of temperature and doping for BaFe$_2$(As$_{1-x}$P$_x$)$_2$ [16]. Black circles and squares indicate $T_s$ (structural transition temperature) and $T_c$, respectively.

With x increase above 0.25, superclusters from AsFe′$_4$ plaquettes are disintegrated and AsFe″$_4$ plaquettes appear, the concentration of which reaches a maximum at ordered arrangement corresponding to x = 0.5 (Fig. 4d). The latter case corresponds to a lower density of CT plaquettes, that should result in lower $T_c$. Therefore, with x increase in the range 0.25 <x <0.5, $T_c$ initially remains constant, as long as the percolation along AsFe′$_4$ plaquettes remains and then decreases to $T_c$(x= 0.5) starting from the site percolation threshold $x_p$ on the lattice $\sqrt{2}\times\sqrt{2}$, that is $x_p$ = 0.593×0.5≈0.3 according to (1).

At x>0.5, some of the plaquettes are under the joint action of three dopants, and the conditions for the appearance of excitons cease to be fulfilled in them. This leads to a rapid decrease in the concentration of CT plaquettes with x increase. Figure 4e shows the doping dependence of $T_c$ for Ba(FeAs$_{1-x}$P$_x$)$_2$ [15]. It is seen that the region of existence of superconductivity coincides with the region of existence of clusters of the CT phase.

It should be noted that, according to the above, the formation of percolation CT clusters implies the ordering of the P-dopant at x = 0.25 and 0.5 into regular lattices. In this case, since the conduction occurs along the CT cluster, the trajectories of the carriers must lie outside the P-sites. This is in agreement with the fact that the P atoms doped directly into the FeAs plane does not induce appreciable scattering [17,18].

As it follows from [19], the residual resistance for single crystals of BaFe$_2$As$_2$ doped with Co, P or K increases approximately proportional to the dopant concentration in the range below the percolation threshold for CT clusters. However, above the percolation threshold, the resistance drops sharply. At that, for Co doping the resistance drops by a factor of ~2, whereas for P or K doping it falls by almost an order. This is in agreement with the fact that the opening of the conduction channel along the CT cluster in all cases leads to a drop in resistance, but in the cases of P and K doping, the drop in resistance is much stronger, since the carrier trajectories in CT cluster lie outside P or K sites. In the same time, it means that in Ba(Fe$_{1-x}$Co$_x$)$_2$As$_2$, Co ions are generally not outside the current paths, i.e. Co ions are not necessarily the centers of ordered trionic complexes. From above we can conclude that in Ba(Fe$_{1-x}$Co$_x$)$_2$As$_2$, trionic complexes are ordered directly against the background of an irregular lattice of dopants.

As follows from the model, the spatial homogeneity of the physical characteristics of the crystals under consideration is determined by the degree of ordering of trionic complexes (in Ba(Fe$_{1-x}$Co$_x$)$_2$As$_2$) or dopants (in BaFe$_2$(As$_{1-x}$P$_x$)$_2$) and by formation of a percolation network of CT plaquettes.

However, at certain concentration values corresponding to the formation of superclusters, the formation of large 2D domains, the skeletons of which are these superclusters, is possible in each FeAs layer. If their sizes are large enough, then they form stacks of 2D domains in the crystal, which can be considered as 3D granules. In this case, the entire crystal at the indicated concentration values can be considered as a granular superconductor in which the granules are connected to each other by Josephson weak links.

How can you verify if this model grasps the structure of a real crystal? Unfortunately, we are not aware of experiments where such a structure in high-quality optimaly doped single crystals would be observed by modern methods of visualization, such as, for example, STS/STM. A possible reason is that these techniques are surface sensitive and more suitable for research of cleavable crystals with uncharged interfaces such as FeSe(Te).

As an indirect confirmation of the granular structure of optimally doped iron pnictide crystals, one can refer to the results of [20], where the authors in the ZFC mode measured the field dependence of the magnetic moment of Ba(Fe$_{0.926}$Co$_{0.074}$)$_2$As$_2$ and Ba$_{0.6}$K$_{0.4}$Fe$_2$As$_2$ single crystals. The linear increase in the magnetic moment with magnetic field was observed up to 5.5 T, that is much higher than both H$_{c1}$ and the upper limit for the surface barrier. Moreover, this behavior was observed both in the "ab" and "c" directions. We believe that it is just this behavior that should be observed in crystals with a grain radius r <$\lambda$, since with a decrease in r below $\lambda$, H$_{c1}$ in them will increase as $(\lambda/r)^2$. The size of these granules can be estimated based on the ratio H$_{c1}\sim\phi_0/r^2$ (here $\phi_0$ is the flux quantum). Since H$_{c1}\sim\phi_0/r^2$>5.5 T, then r <20 nm. For the detecting of charge ordering in granules with r ~ 10 - 20 nm, the RXS technique (resonant X-ray scattering) could be recommended, that has been successfully used to detect charge-density modulations in cuprates with a spatial coherence as short as 1.5–2 nm [21]

**The doping dependences of the London penetration depth in Ba(Fe$_{1-x}$Co$_x$)$_2$As$_2$ and BaFe$_2$(As$_{1-x}$P$_x$)$_2$**

Additional support of the validity of the proposed model can be obtained from an analysis of the results of measurements of the dependence of the London penetration depth $\lambda$ on the doping level in Ba(Fe$_{1-x}$Co$_x$)$_2$As$_2$ and BaFe$_2$(As$_{1-x}$P$_x$)$_2$ [8,22,23]. As can be seen from (Fig. 6), for both compounds there is sharp increase in $\lambda$ in the region of optimal doping.

Usually, when considering the results of measurements of the London penetration depth, it is assumed that the superconducting medium is homogeneous, and $\lambda$ is determined entirely by the density of the superconducting pairs. In such a picture, the appearance of a sharp peak of $\lambda$ in a narrow doping range does not find a satisfactory explanation.

In [24-27], it was suggested that there is a quantum phase transition (QPT) in pnictides at a certain doping level. On the base of this assumption, the appearance of a certain anomaly in the behavior of $\lambda(x)$ in the vicinity of QPT, due to quantum fluctuations, was predicted. However, there is no consensus regarding the shape of this anomaly. At the same time, the assumption that the peak of $\lambda$ is associated with QPT does explain neither its amplitude nor the fact that this peak is located between two close minima (as in Ba(Fe$_{1-x}$Co$_x$)$_2$As$_2$), and its width is much smaller than the region of coexistence of superconductivity and antiferromagenetism.

We offer the different explanation for the anomalous λ(x) dependence. As follows from the above consideration, Ba(Fe$_{1-x}$Co$_x$)$_2$As$_2$ and BaFe$_2$(As$_{1-x}$P$_x$)$_2$ can actually be simulated in a certain concentration range as granular superconductor. For the analysis of magnetic field penetration in this range it can be considered as Josephson media, in which the superconducting domains are weakly coupled. In granular Josephson media, λ is defined as $λ^2 \propto 1/g$, where g≈E$_j$d is the "rigidity" of the medium, E$_j$ is the energy of Josephson bonds per unit area, and d is the size of the domain [28]. With the decay of large domains into small ones, the rigidity of the superconducting medium will decrease and λ, accordingly, will increase. This is what we believe is observed in [8,22,23].

In the case of Ba(Fe$_{1-x}$Co$_x$)$_2$As$_2$ large 3D superconducting domains can be formed only upon ordering of Co ions in sublattices with $l=\sqrt{18}$ and $l=4$, that takes place at x=0.056 and x=0.062. The same concentration values correspond to the maxima of the rigidity of the Josephson medium and, accordingly, to the minima of the London penetration depth. In the intermediate interval 0.056<x<0.062 the London penetration depth will have a maximum. The indicated position of the maximum is in good agreement with the experimental result [8].

It is also possible to estimate the expected peak amplitude. If at a concentration corresponding to the first minimum there are clusters of only one type, and at a concentration corresponding to the second minimum, there are those of a different type, then at an intermediate concentration there are clusters of both types, and, consequently, their size is two times smaller, and the value of λ at this point, according to the theory, is $\sqrt{2}$ times more. This estimate is in good agreement with the result of the experiment [8].

In this regard, it is important to mention the work [29] where the authors observed the coexistence of frozen antiferromagnetic domains and superconductivity in optimally doped Ba(Fe$_{1-x}$Co$_x$)$_2$As$_2$ using the NMR spectroscopy. In this case, the characteristic sizes of antiferromagnetic domains change by more than an order of magnitude, reaching the maximum inhomogeneity at x ≈ 0.06. This result should be considered as a confirmation of the proposed model, according to which, upon decay of large superconducting domains in a narrow region of x, many small superconducting and antiferromagnetic domains coexist.

As for isovalently doped BaFe$_2$(As$_{1-x}$P$_x$)$_2$, the mechanism of the appearance of the peak of London penetration depth has the same nature as in Ba(Fe$_{1-x}$Co$_x$)$_2$As$_2$ considered above. In this case, λ has minima at x=0.25 and 0.5, which correspond to an ordered arrangement of dopants with the possibility of forming superclusters (Fig. 4 c, d). With an increase in x>0.25, large clusters disintegrate, which is accompanied by a sharp increase in λ. This growth stops at x=0.3, which corresponds to the site percolation threshold $x_p = 0.593/l^2$ (1) for clusters with an ordering in the $\sqrt{2}x\sqrt{2}$ lattice (Fig. 4d). At x>0.3, clusters with this type of ordering begin to grow.

Thus, the above consideration allows us to assert that the nature of the sharp peaks in the dependence of the London penetration depth on doping is associated with the disintegration of large superconducting domains in a certain concentration range.

As noted, x values, corresponding to the minima of λ, correspond to the formation at these x of superconducting superclusters of maximum size, but of different types of ordering. To the left and to the right of these values, λ will increase as the sizes of superconducting superclusters decrease. This increase can be estimated from the ratio of the power of percolation cluster just above the percolation threshold $x_p$ to that at complete ordering. Since at the percolation threshold the percolation cluster includes about half of all sites, then λ at $x_p$ should increase $\sqrt{2}$ times in comparison with λ at complete ordering. In Ba(Fe$_{1-x}$Co$_x$)$_2$As$_2$ for cluster with $l=\sqrt{18}$ $x_p$≈0.033 (1). Therefore, the ratio λ(x=0.056)/λ(x=0.033)≈$\sqrt{2}$ must be satisfied. Since the experimental measurements of λ start from x≈0.4, we take for estimation that in the range 0.033 <x <0.04 the power of the percolation cluster changes insignificantly and λ (x=0.33)≈λ(x=0.4). Measurements give λ(x=0.056)/λ(x=0.04)≈1.5, which is in good agreement with our estimate. For $BaFe_2(As_{1-x}P_x)_2$, such an estimate is impossible, since with doping decrease the crystal, before reaching the percolation threshold (x ≈ 0.15), passes at x<0.2 into the non-superconducting SDW state.

As for the existence of a similar peak in $Ba(Fe_{1-x}Ni_x)_2As_2$, there are two concentration values at which the formation of large superconducting domains is possible: x=1/36 and x=1/32. Therefore the peak of λ should be sought in the range 1/36<x<1/32 (0.028<x<0.031), i.e. where the resistance anisotropy (Fig. 2) has a sharp minimum. However, due to small width of this range, the observation the peak of λ in this compound is a difficult task.

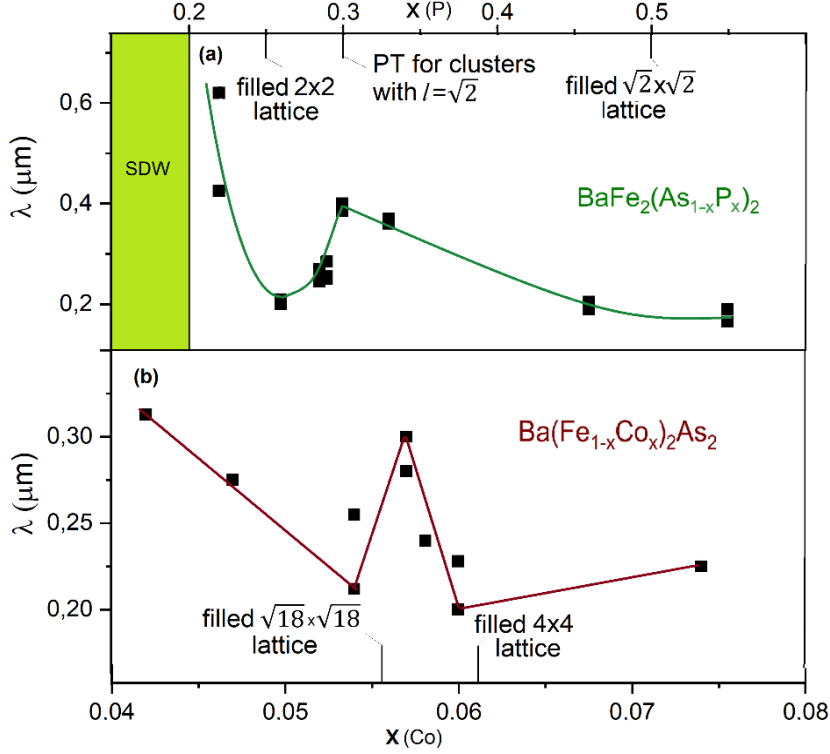

**Figure 6.** The dependences of the London penetration depth λ on the doping level in $Ba(Fe_{1-x}Co_x)_2As_2$ and $BaFe_2(As_{1-x}P_x)_2$ according to the data of works [8] and [23], respectively. PT - percolation threshold.

**Conclusion**

Thus, the proposed model, despite its simplicity and neglect of the features of the electronic structure, makes it possible not only to accurately calculate the positions of superconducting domes on the phase diagrams of specific HTSC compounds, but also to explain the nature and position of sharp peaks in the London penetration depth and non-monotonic dependence of resistivity anisotropy depending on the doping level. Taking into account the capabilities of the model in terms of accurately determining of the position of various features on the phase diagrams of cuprates and iron pnictides, it can be assumed that the assumptions on the local character of doping and the cluster nature of the superconducting phase are quite justified. In turn, this can be regarded as confirmation that high-temperature superconductivity arises in clusters of CT plaquettes, where band electrons can interact with pair states of HL centers. Considering that the interaction of itinerant electrons with excitonic states is genetically inherent in proposed model, it can be assumed that the exciton mechanism contributes to superconducting pairing in cuprates and pnictides.

**CRediT authorship contribution statement**

**K. Mitsen**:. Conceptualization, Model construction, Writing – original draft. **O. Ivanenko**. Development of this work, Supervision, Writing – review & editing.

**Declaration of competing interest**

The authors declare that they have no known competing financial interests or personal relationships that could have appeared to influence the work reported in this paper.

**Acknowledgments**

The research was carried out within the state assignment of Ministry of Science and Higher Education of the Russian Federation (theme No. AAAA-A19-119083090048-5).

**References**

1. McElroy K, Lee J; Slezak JA, Lee D-H, Eisaki H; Uchida S, et al. Atomic-Scale Sources and Mechanism of Nanoscale Electronic Disorder in $Bi_2Sr_2CaCu_2O_{8+d}$. Science 2005;309(5737):1048-52.
2. Gavrichkov VA, Shan'ko Y, Zamkova NG, Bianconi A. Is there any hidden symmetry in the stripe structure of perovskite high-temperature superconductors? J. Phys. Chem. Lett. 2019;10(8):1840-44.
3. Kresin VZ, Wolf SA. Inhomogeneous Superconducting State and Intrinsic $T_c$: Near Room Temperature Superconductivity in the Cuprates. J. Supercond. Nov. Magn. 2012;25(2):175-80.
4. Mitsen KV, Ivanenko OM. Superconducting phase diagrams of cuprates and pnictides as a key to understanding the HTSC mechanism. Phys. Usp. 2017;60(4):402-11.
5. Mitsen KV, Ivanenko OM. Phase diagram of $La_{2-x}M_xCuO_4$ as the key to understanding the nature of high-Tc superconductors. Physics-Uspekhi 2004;47(5);493-510
6. Mitsen K, Ivanenko O. Towards the issue of the origin of Fermi surface, pseudogaps and Fermi arcs in cuprate HTSCs. J. Alloys Compd. 2019;791:30–8.
7. Li LJ, Luo YK, Wang QB, Chen H, Ren Z, Tao Q, et al. Superconductivity induced by Ni doping in $BaFe_2As_2$ single crystals. New J. Phys. 2009;11(2):025008
8. Joshi KR, Nusran NM, Tanatar MA, Cho K, Bud'ko SL, Canfield PC, et al. Quantum phase transition inside the superconducting dome of $Ba(Fe_{1-x}Co_x)_2As_2$ from diamond-based optical magnetometry. New J. Phys. 2020;22(5):053037.
9. Jacobsen, J.L. High-precision percolation thresholds and Potts-model critical manifolds from graph polynomials. J. Phys. A: Math. Theor. 2014;47(13):135001
10. Ishida S, Nakajima M, Liang T, Kihou K, Lee C-H, Iyo A, et al. Effect of Doping on the Magnetostructural Ordered Phase of Iron Arsenides: A Comparative Study of the Resistivity Anisotropy in Doped BaFe2As2 with Doping into Three Different Sites. J. Am. Chem. Soc. 2013;135(8):3158−63
11. Axe J, Moudden A, Hohlwein D, Cox D, Mohanty K, Moodenbaugh A, et al. Structural Phase Transformations and Superconductivity in $La_{2-x}Ba_xCuO_4$. Phys. Rev. Lett. 1989;62(23):2751-54.
12. Nandi S, Kim M., Kreyssig A, Fernandes RM, Pratt DK, Thaler A, et al. Anomalous Suppression of the Orthorhombic Lattice Distortion in Superconducting $Ba(Fe_{1-x}Co_x)_2As_2$ Single Crystals. Phys. Rev. Lett. 2010;104(5):057006
13. Bud'ko SL, Ni Ni, Canfield PC. Jump in specific heat at the superconducting transition temperature in $Ba(Fe_{1-x}Co_x)_2As_2$ and $Ba(Fe_{1-x}Ni_x)_2As_2$ single crystals. Phys. Rev.B 2009;79(22):220516

14. Kuo HH, Chu JH, Scott CR, Yu L, McMahon PL, Kristiaan DG, et al. Possible origin of the nonmonotonic doping dependence of the in-plane resistivity anisotropy of Ba(Fe$_{1-x}$T$_x$)$_2$As$_2$ (T = Co, Ni and Cu). Phys. Rev. B;84(5):054540.
15. Kasahara S, Shibauchi T, Hashimoto K, Ikada K, Tonegawa S, Okazaki R, et al. Evolution from non-Fermi- to Fermi-liquid transport via isovalent doping in BaFe$_2$(As$_{1-x}$P$_x$)$_2$ superconductors. Phys. Rev. B 2010;81(18):184519
16. Kuo, H.H. (2014). Electronic nematicity in iron-based superconductors (PhD Theses, Stanford University, USA) https://purl.stanford.edu/gd993cz5400.
17. Shishido H, Bangura AF, Coldea AI, Tonegawa S, Hashimoto K, Kasahara S, et al. Evolution of the Fermi surface of BaFe$_2$(As$_{1-x}$P$_x$)$_2$ on entering the superconducting dome. Phys. Rev. Lett. 2010;104(5):057008.
18. van der Beek CJ, Konczykowski M, Kasahara S, Terashima T, Okazaki R, Shibauchi T, Matsuda Y. Quasiparticle scattering induced by charge doping of iron-pnictide superconductors probed by collective vortex pinning. Phys. Rev. Lett. 2010;105(26):267002.
19. Ishida S, Nakajima M, Liang T, Kihou K, Lee C-H, Iyo A, et al. Effect of Doping on the Magnetostructural Ordered Phase of Iron Arsenides: A Comparative Study of the Resistivity Anisotropy in Doped BaFe2As2 with Doping into Three Different Sites. J. Am. Chem. Soc. 2013;135(8):3158−63
20. Prozorov R, Tanatar MA, Shen B, Cheng P, Wen H-H, Bud'ko SL, Canfield P. Anomalous Meissner effect in pnictide superconductors. Phys. Rev. B 2010;82(18);180513(R)
21. Comin R, Damascelli A. Resonant X-Ray Scattering Studies of Charge Order in Cuprates. Annu. Rev. Condens. Matter Phys. 2016;7:369-405.
22. Hashimoto K. Cho K, Shibauchi T, Kasahara S, Mizukami Y, Katsumata R, et al. A sharp peak of the zero-temperature penetration depth at optimal composition in BaFe$_2$(As$_{1-x}$P$_x$)$_2$. Science 2012;336(6088):1554-7.
23. Lamhot Y, Yagil A, Shapira N, Kasahara S, Watashige T, Shibauchi T, et al. Local characterization of superconductivity in BaFe$_2$(As$_{1-x}$P$_x$)$_2$. Phys. Rev. B 2015;91(6):060504(R).
24. Chowdhury D, Swingle B, Berg E, Sachdev S. Singularity of the London penetration depth at quantum critical points in superconductors. *Phys. Rev. Lett.* 2013;111(15):157004
25. Levchenko A, Vavilov MG, Khodas M, and Chubukov AV. Enhancement of the London Penetration Depth in Pnictides at the Onset of Spin-Density-Wave Order under Superconducting Dome. *Phys. Rev. Lett.* 2013;110(17):177003
26. Nomoto T, Ikeda H. Effect of Magnetic Criticality and Fermi-Surface Topology on the Magnetic Penetration Depth. *Phys. Rev. Lett.* 2013;111(16):167001.
27. Chowdhury D, Orenstein J, Sachdev S and Senthil T. Phase transition beneath the superconducting dome in BaFe$_2$(As$_{1-x}$P$_x$)$_2$. *Phys. Rev.* B 2015;92(08):081113
28. Sonin EB, Tagantsev AK. Electrodynamics of the Josephson medium in high-T$_c$ superconductors. Phys. Lett. A 1989;140(3):127-32.
29. Dioguardi AP, Crocker J, Shockley AC, Lin CH, Shirer KR, Nisson DM, et al. Coexistence of Cluster Spin Glass and Superconductivity in Ba(Fe$_{1-x}$Co$_x$)$_2$As$_2$ for 0.060≤x≤0.071. Phys. Rev. Lett. 2013;111(20):207201.